\documentclass[preprint]{aastex}
\usepackage{ulem}

\shorttitle{Twin clusters at $z=1.62$}
\shortauthors{Tanaka et al.}

\begin{document}


\title{The spectroscopically confirmed X-ray cluster at $z=1.62$ with
a possible companion in the Subaru/XMM-Newton deep field}


\author{Masayuki Tanaka}
\affil{Institute for the Physics and Mathematics of the Universe, The University of Tokyo,  5-1-5 Kashiwanoha, Kashiwa-shi, Chiba 277-8583, Japan}
\affil{European Southern Observatory, Karl-Schwarzschild-Str. 2, D-85748 Garching bei M\"{u}nchen, Germany}

\author{Alexis Finoguenov}
\affil{Max-Planck-Institut f\"{u}r extraterrestrische Physik, Giessenbachstrasse, D-8574 Garching bei M\"{u}nchen, Germany}
\affil{University of Maryland, Baltimore County, 1000 Hilltop Circle,  Baltimore, MD 21250, USA}

\and

\author{Yoshihiro Ueda}
\affil{Department of Astronomy, Kyoto University, Kyoto 606-8502, Japan}




\begin{abstract}
We report on a confirmed galaxy cluster at $z=1.62$.
We discovered two concentrations of galaxies at $z\sim1.6$
in the Subaru/XMM-Newton deep field based on deep multi-band
photometric data.
We made a near-IR spectroscopic follow-up observation of
them and confirmed several massive galaxies at $z=1.62$.
One of the two is associated with an extended X-ray emission 
at $4.5\sigma$ on a scale of $0'.5$, which is typical of high-$z$ clusters.
The X-ray detection suggests that it is a gravitationally bound system.
The other one shows a hint of an X-ray signal, but only at $1.5\sigma$,
and we obtained only one secure redshift at $z=1.62$.
We are not yet sure if this is a collapsed system.
The possible twins exhibit a clear red sequence at $K<22$
and seem to host relatively few number of faint red galaxies.
Massive red galaxies are likely old galaxies -- they have colors
consistent with the formation redshift of $z_f=3$ and a spectral
fit of the brightest confirmed member yields an age of
$1.8_{-0.2}^{+0.1}$ Gyr with a mass of $2.5_{-0.1}^{+0.2}\times10^{11}\rm M_\odot$.
Our results show that it is feasible to detect clusters
at $z>1.5$ in X-rays and also to perform detailed
analysis of galaxies in them with the existing
near-IR facilities on large telescopes.
\end{abstract}




\section{Introduction}

The galaxy evolution is closely linked to the structure formation
of the Universe.
Isolated galaxies and those in clusters evolve in different ways
and the differential evolution results in the strong environmental
dependence of galaxy properties observed in the local Universe.
The environmental dependence of galaxy properties is already
strong at $z=1$.  Intensive studies of $z\sim1$ clusters have shown
that red galaxies have already become the dominant population
in clusters at $z=1$ (e.g., \citealt{blakeslee03,nakata05,lidman08,mei09}).
The spectroscopically confirmed highest redshift X-ray cluster known until now
is located at $z=1.45$ \citep{stanford06}, but 
red galaxies are abundant even in this highest-$z$ cluster \citep{hilton09}.
A few authors reported a possible clusters at $z>1.5$.
For example, \citet{kurk09} presented an over-density of galaxies at $z=1.6$,
which may later collapse to a cluster.
It is not detected in X-rays down to a limit of
 $<1\times10^{-16}\rm ergs\ s^{-1}\ cm^{-2}$.
\citet{andreon09} presented a cluster at $z_{phot}\sim1.9$, but
its redshift has been questioned by Bielby et al. (in prep) ---
it is likely a complex system with one confirmed at $z=1.1$
and another one possibly at $z\sim1.5$ (see Bielby et al. for details).
It is challenging to identify clusters at $z>1.5$, but
one has to explore this redshift regime to identify the epoch
when red galaxies become the dominant population in clusters and
how the environmental dependence of galaxy properties is established.

We are conducting a distant X-ray cluster survey in
the Subaru/XMM-Newton Deep Field (SXDF).
We refer the reader to \citet{finoguenov09b} for details of our survey
and the construction of the deep multi-band photometric catalog\footnote{
We note that we have revised the X-ray catalog presented in \citet{finoguenov09b}
by adding the 0.3-0.5 keV band data from XMM.
We extract the images separately using the single events,
produce a corresponding background estimation, and
add them to the mosaic only at the very last stage.
This increases signals from distant low-mass clusters.
}.
As part of the survey, we identified two concentrations of red galaxies
first by their red sequence and then we found one of the sources is
also securely detected in X-rays.
We estimated their redshifts to be $z\sim1.6$ both from the location
of the cluster red sequence and from photometric redshifts.
Not only they have very similar redshifts, but they are also close to
each other on the sky; they are separated only by $\sim2.5$ arcmin
($\sim1.3$ Mpc at $z=1.6$).
Followed by the initial photometric identification, we carried out
a near-IR spectroscopic follow-up observation of the possible twin clusters
and we report on the results in the paper.
Recently, \citet{papovich10} presented a completely independent
study of one of the clusters.
Readers are referred to their paper for a similar, but independent
analysis.

The layout of the paper is as follows.  We first describe the near-infrared
spectroscopic follow-up observation and present a discovery of the most
distant X-ray cluster from the observation in section 2.
Section 3 discusses our results and summarizes the paper.
Unless otherwise stated, we adopt H$_0=72\rm km\ s^{-1}\ Mpc^{-1}$,
$\Omega_{\rm M}=0.25$, and $\Omega_\Lambda =0.75$.
Magnitudes are on the AB system.

\section{Observation and Discovery}

We carried out near-IR multi-object spectroscopy with MOIRCS
on the Subaru telescope \citep{ichikawa06,suzuki08} on the 3rd-4th of December 2009.
We selected targets for spectroscopy using photometric redshifts ;
the first priority is given to red galaxies at $1.4<z_{phot}<1.8$,
and second priority to blue galaxies at the same redshift range.
The observing conditions were poor; $1-1.2$ arcsec seeing
in the $J$-band most of the time and we occasionally had
clouds and high humidity. 
We used the $zJ500$ grism with a slit width of 0.8 arcsec,
which gave a resolving power of $R\sim500$.
The exposures were broken into 15 min. to 20 min. each and the total
integration time amounted to 320 min.
The data were reduced in a standard manner with custom-designed pipeline \citep{tanaka09}.
The spectra were visually inspected and redshifts and confidence
flags were assigned to each object.
We obtained 11 secure redshifts and 5 possible redshifts
out of 39 observed objects.
We present sample spectra in Fig. \ref{fig:spec}.


Fig. \ref{fig:twin_clusters} summarizes our effort.
There are two distinct concentrations of red galaxies.
The concentration on the top-right (SXDF-XCLJ0218-0510) is detected in X-rays
at $4.5\sigma$, which suggests that it is a gravitationally bound system.
The X-ray is extended towards West, but we suspect that this is because
a gap between the pn detectors passes the Eastern part, where only MOS data
contributed for the detection.
It has two very bright galaxies and they have secure redshifts at
$z=1.625$ and $z=1.634$, respectively.
There are two more galaxies with secure redshifts at $z\sim1.625$
in this system.
Together with the clear concentration of the photo-z selected galaxies,
we are highly confident that this system is a real cluster.
This is the first confirmation of a distant cluster based on near-IR spectroscopy.

\citet{papovich10} confirmed several low-mass blue galaxies
at the same redshift, one of which is a common object at $z=1.649$.
Including redshifts from \citet{papovich10}, we have 9 secure redshifts
in this system, and we measure its redshift to be
$z=1.6230_{-0.0003}^{+0.0009}$ and a velocity dispersion of
$537_{-213}^{+76}\ \rm km\ s^{-1}$
using the biweight estimator and gapper method, respectively \citep{beers90}.
{We note that biweight estimator gives a velocity dispersion of
$109_{-30}^{+350}\ \rm km\ s^{-1}$, being consistent with the one from gapper
within the errors.
We take the one from gapper in the following because of its robustness against
poor statistics \citep{beers90}.
The virial mass and radius are $\rm M_{200}=1.1_{-0.8}^{+0.5}\times10^{14}\rm M_\odot$
and $R_{200}=0.54_{-0.21}^{+0.08}$ Mpc \citep{carlberg97}.
We note that one obtains a biased high velocity dispersion when
only blue galaxies are used \citep{biviano06}, but half of the galaxies
used here are massive red galaxies.
Two galaxies at $z=1.642$ and $z=1.649$ might be background
galaxies because they increase the velocity dispersion to
$1183_{-389}^{+103} \ \rm km\ s^{-1}$, which seems too large,
but we are not sure at this point.

The properties of the system derived from X-rays are consistent
from those from optical. 
We detect an apparent flux of $8.1\pm3.7\times10^{-16}\rm\ ergs\ s^{-1}\ cm^{-2}$
at 0.5-2.0 keV (37 counts) after the point source removal \citep{finoguenov09b}, which can be
translated into $L_X=3.4\pm1.6\times10^{43}\rm\ ergs\ s^{-1}$ at $z=1.6$
in the rest-frame 0.1-2.4 keV.
Then, assuming the relation derived by \citet{leauthaud10}, we obtain
a cluster mass of $M_{200}=5.7\pm1.4\times10^{13}\rm M_\odot$
and a virial radius of $R_{200}=0.440$ Mpc, being consistent with the optical estimates.
Cluster temperature is estimated to be $T_X=1.7\pm0.3$ keV
from the $L_X-T_X$ relation as mentioned in \citet{leauthaud10}.

We emphasize that most previous cluster studies did not remove
fluxes from unresolved point sources (AGNs) once the extent of
a cluster has been established.
This procedure no longer works in deep surveys because
the contamination from point sources is significant.
We have therefore adopted the procedure in which we always remove
the flux expected from detected point sources based on the shape
of the PSF.  In fact, we have attributed a large fraction ($\sim50\%$)
of the emission around the cluster to point sources.
At the XMM exposure around the cluster,
if a point source contributes substantially to the overall
emission ($>30$\%), it can be detected individually.
We can therefore exclude a possibility of large contamination
from undetected point sources to the measured extended flux.
However, some of the detected point sources might be associated
with the cluster core and high resolution X-ray imaging will allow us
to identify potential confusions.

The one on the bottom-left (SXDF-XCLJ0218-0512) in Fig. \ref{fig:twin_clusters}
hosts an outstandingly bright galaxy, which is likely a cD galaxy of this system.
We obtain a possible redshift of $z=1.625$ for this galaxy.
Interestingly, we  observe a hint of an X-ray emission right on
top of the cD galaxy at $1.5\sigma$.
Note that we do not observe nebular emission lines from the cD galaxy
(see the bottom panel in Fig. \ref{fig:spec}),  but we need deeper
observations to confirm if the X-ray signal is real and if
it is from an AGN or the cluster core.
We confirm another galaxy at $z=1.627$ nearby.
Although the concentration of galaxies around the possible cD galaxy
is clear, we are not yet sure if this is a collapsed system due to
the weakness of X-ray emission and to too few spectroscopic redshifts.
For the rest of the paper, we assume that the two systems are real clusters
at the same redshift and combine them to gain statistics.
We note that our results remain unchanged if we use the secure cluster only.

\section{Results and Discussion}

The cluster red sequence is a ubiquitous feature of galaxy clusters.
We draw color-magnitude diagrams in Fig. \ref{fig:cmd} using galaxies
at $<1$ Mpc from the twins.
We refer the readers to \citet{papovich10} for a similar analysis.
The photo-$z$ selected galaxies form a clear red sequence at $K<22$,
while the sequence is not very clear at fainter magnitudes.
We model the location of the red sequence with the updated version
of the \citet{bruzual03} code, which takes into account effects of
thermally pulsating AGB stars, using the procedure described in \citet{lidman08}.
Here we assume the Chabrier IMF \citep{chabrier03}.
A biweight fit to the red galaxies gives a sequence very close
to the $z_f=3$ model sequence, suggesting that they are relatively old galaxies.
The cosmic time between $z=3$ and 1.6 is 2.0 Gyr.
We note that we obtain $z_f=3$ or higher if we use the $z-J$ color as used
in \citet{papovich10} who obtained $z_f=2.4$.
We do not know the cause of the difference in $z_f$ at this point, but
it could be due to the rather old UKIDSS catalog they used.
Their paper is still under review at the time of this writing and
we do not pursue this point further.
The $z-K$ color is more sensitive to $z_f$ at this redshift\footnote{
The $z-K$ color of the model red sequence formed at $z_f=5$
is redder than the $z_f=2$ sequence by 0.74,
while the difference is only 0.26 for the $z-J$ color.
Thus, the $z-K$ color is more sensitive to $z_f$ than $z-J$.
} and we prefer to use it in this paper.
As a sanity check, we perform a statistical subtraction of
fore-/background galaxies without using photo-$z$
following the recipe by \citet{tanaka05} in the right panel.
A clump of red galaxies is clearly visible in the right panel as well.
Again, the sequence vanishes at $K>22$.
There are a number of proto-clusters at even higher redshift that have
a confirmed over-density of low-mass star forming galaxies \citep{miley08},
but Fig \ref{fig:cmd} shows that a near-IR spectroscopic observation is essential to
confirm the dominant population of massive galaxies in $z>1.5$ clusters.

To further quantify the red sequence, we plot luminosity functions
(LFs) of red galaxies in Fig. \ref{fig:lf}.
Here we define red galaxies as those having $\Delta |z-K|<0.2$
from the best fitting red sequence.
We confirmed that a small change in the definition does not alter
the result below.
But, we should note that the current statistics is very poor.
We use galaxies at $1.4<z_{phot}<1.8$
and statistically subtract the remaining contamination
using the entire SXDF field as a control field sample.
We compare the $z=1.6$ LF with those at lower redshifts
taken from \citet{tanaka08}.
We plot LFs relative to $m_K^*$ at each redshift to correct for
the passive evolution.
The LF of the red galaxies in the twins is similar to the group LF at $z=1.1$,
which suggests a deficit of faint red galaxies.  On the other hand,
it seems that clusters at $z=1.6$ already host abundant massive galaxies.
As seen in Fig. \ref{fig:cmd}, these massive galaxies are mostly red galaxies
and the clusters host few bright blue galaxies.
It is hard to generalize the trend we see here with only a few clusters, but
the implication  would be that massive red galaxies might have
become the dominant population in clusters already at $z=1.6$.

Finally, we present a spectral analysis of the brightest
red galaxy with secure redshift in Fig. \ref{fig:specfit}.
We fit the observed spectra with model spectra generated by
the updated version of the \citet{bruzual03} code.  We assume the exponentially
decaying star formation histories and use age, dust extinction,
star formation time scale, and metallicity as free parameters.
We adopt the Chabrier IMF \citep{chabrier03}\footnote{
If we use the Salpeter IMF, we obtain 1.8 times higher stellar mass.
The other properties are nearly the same.
} and fix the redshift at $z_{spec}$.
The fit gives an age of $1.8_{-0.2}^{+0.1}$ Gyr and a small extinction of
$A_V=0.2_{-0.1}^{+0.2}$ mag.
Its star formation time scale is rather short, $0.1_{-0.1}^{+0.2}$ Gyr.
Assuming the exponential decay, it assembled 80\% of its stars at
$z_f\sim2.5$ that it would have at $z=0$, which is in agreement with
that inferred from the $z-K$ color.
Note that $z_f=3$ obtained from the red sequence gives an average
formation redshift of the red galaxies, while we are specifically
discussing the brightest red galaxy with secure redshift here.
The current star formation rate is $<0.1\rm M_\odot\ yr^{-1}$,
which is consistent with the absence of any emission lines
in the observed spectrum.
Its mass is estimated to be $2.5_{-0.1}^{+0.2}\times10^{11}\rm  M_\odot$,
revealing the presence of such a massive galaxy in a $z=1.6$ cluster.
As seen in lower redshift clusters, massive cluster galaxies form early
and evolve passively  (e.g., \citealt{thomas05}).

This study is the first result on cluster galaxies at $z>1.5$
using near-IR spectrograph.
Our observation suggests that $z=1.6$ clusters exhibit a red sequence
at bright magnitudes and are likely luminous in X-rays, which may have
an implication for future high-$z$ cluster surveys.
The cluster we discovered will evolve to a 4--5 keV cluster at $z=0$ \citep{vandenbosch02}.
Such clusters are a sensitive probe of cosmology at high redshifts and
yet its estimated temperature preclude a detection in a Sunyaev-Zeldovich observation.
This demonstrates the power of X-ray observations in finding distant clusters.
The redshift of $z>1.5$ is the epoch when early dark energy is
important and large-scale structures at $z>1.5$ are a robust
tracer of the primordial local non-Gaussianity
(e.g., \citealt{bartelmann06,sadeh07,grossi09}).
More massive clusters than the one reported in this paper may be found
in shallower X-ray observations down to $10^{-14}\rm ergs\ s^{-1}\ cm^{-2}$,
but one needs to survey an order of $5,000$ square degrees \citep{finoguenov09b},
which may be difficult to follow-up photometrically.
However, a similar yield in the number of clusters can be achieved by
reaching $10^{-15}\rm ergs\ s^{-1}\ cm^{-2}$ over 100 square degrees.
Therefore, deep X-ray surveys at high spatial resolution have
a unique window for cosmological studies at the 
time when Universe was only a third of its present age.
Our observation also shows that detailed analysis of galaxy populations
at this redshift regime is feasible with the current near-IR facilities.
This is an encouraging result and motivates us to push
our X-ray cluster survey forward.

\acknowledgments

This study is based on data taken at the Subaru Telescope
operated by the National Astronomical Observatory of Japan
through program S09B-042.
We thank Ichi Tanaka for his support during the observation.
We thank the anonymous referees for their comments, which helped
improve the paper.





\clearpage

\begin{figure}
\epsscale{1.0}
\plotone{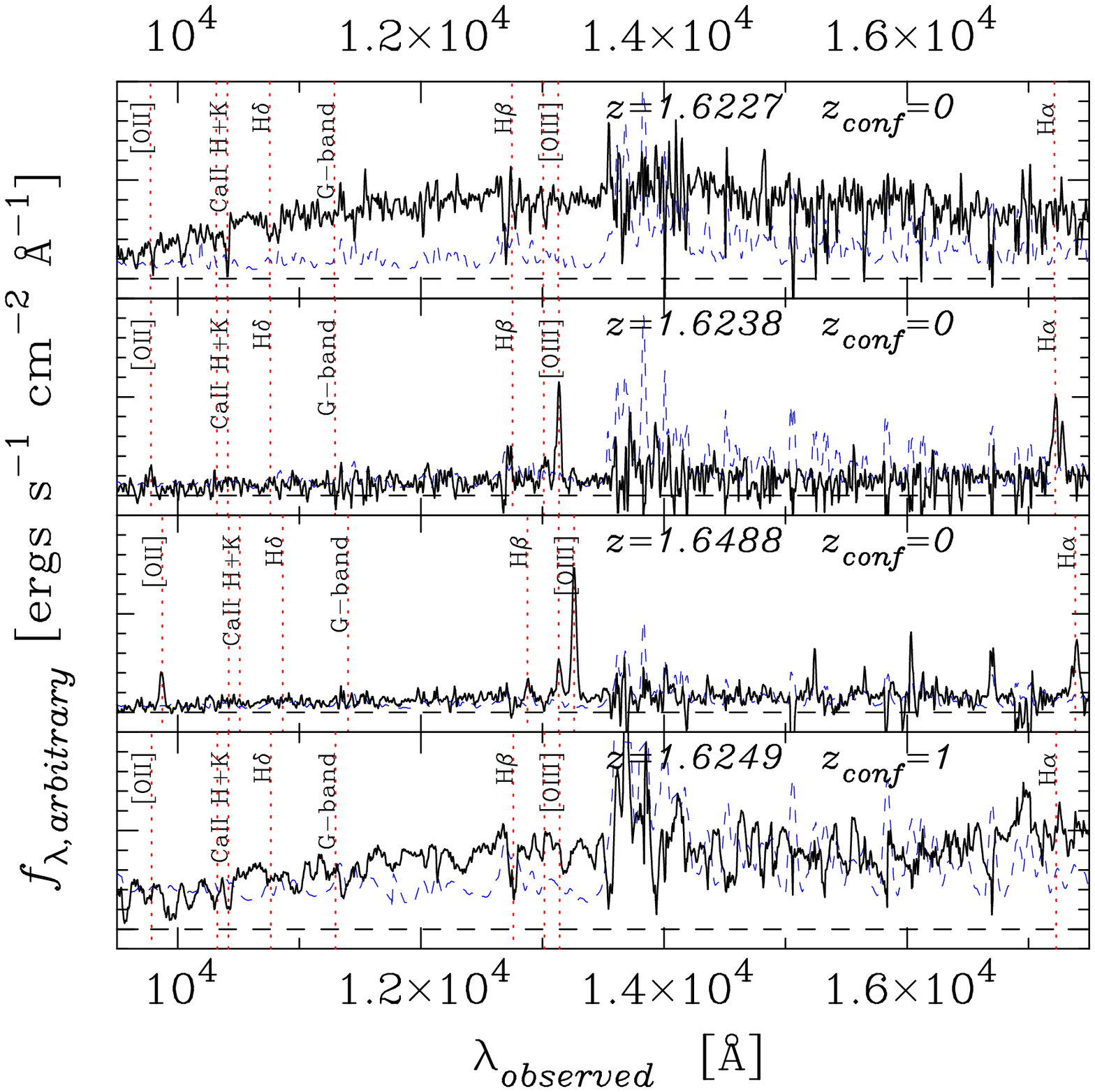}
\caption{
Sample spectra of $z=1.6$ galaxies.
The dashed lines are the noise spectra.
Some of the most prominent spectral features are indicated with the dotted lines.
$z_{conf}=0$ and 1 mean secure and possible redshifts, respectively.
The bottom spectrum is smoothed over 5 pixels.
}
\label{fig:spec}
\end{figure}

\clearpage

\begin{figure}
\epsscale{1.0}
\plotone{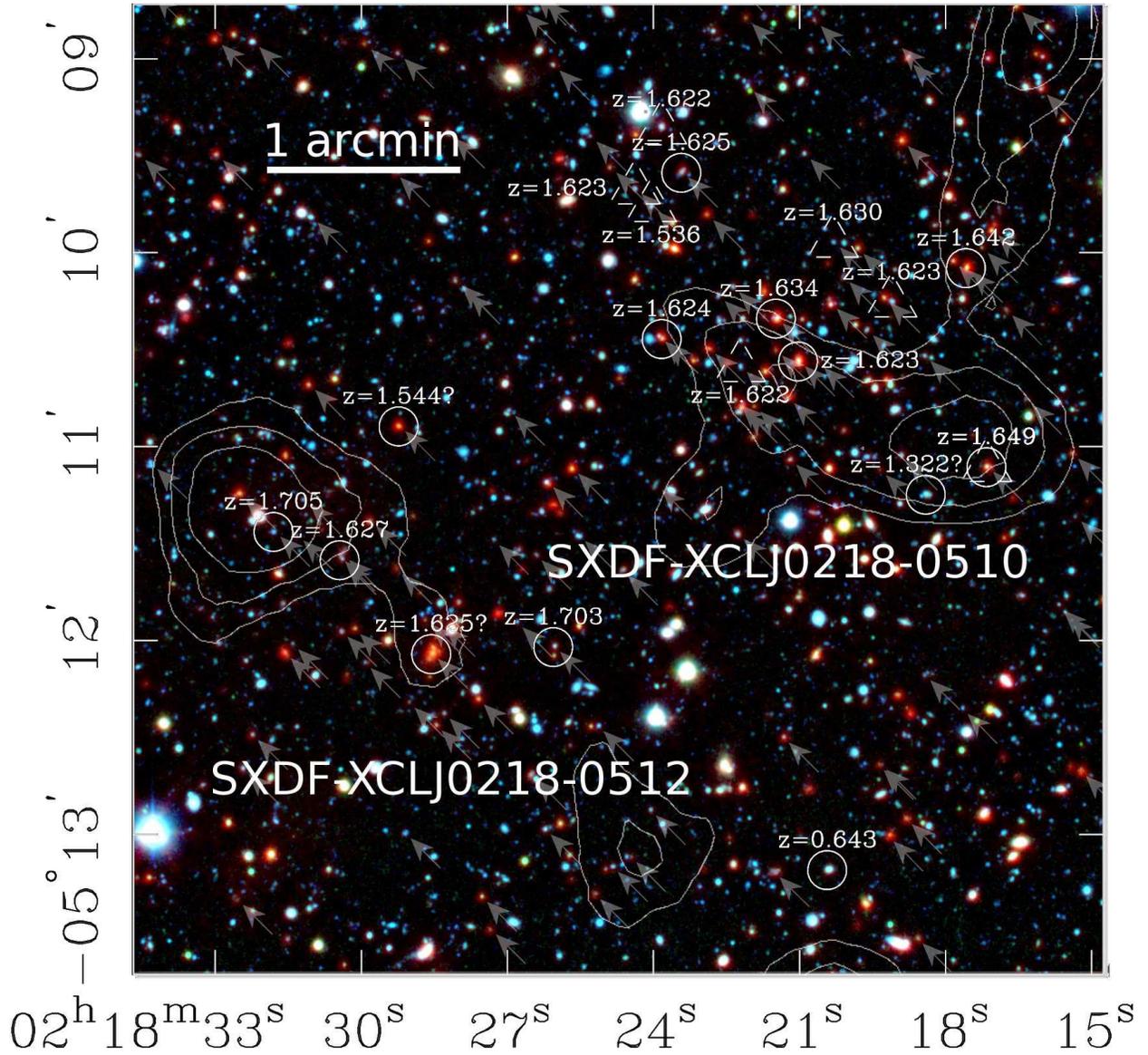}
\caption{
$Rz3.6\mu m$ pseudo-color image of the twins.  The image is 5 arcmin on a side.
The contours show diffuse X-ray emission at the $1.5\sigma$, $3\sigma$, and
$4.5\sigma$ levels in the 0.3--2.0 keV band and point sources are removed.
The arrows indicate photo-$z$ selected potential members ($1.4<z_{phot}<1.8$).
The circles show our objects with spectroscopic redshifts and
the triangles indicate spectroscopic objects from \citet{papovich10}.
The redshifts with '?' are possible redshifts.
}
\label{fig:twin_clusters}
\end{figure}

\clearpage

\begin{figure}
\epsscale{1.0}
\plotone{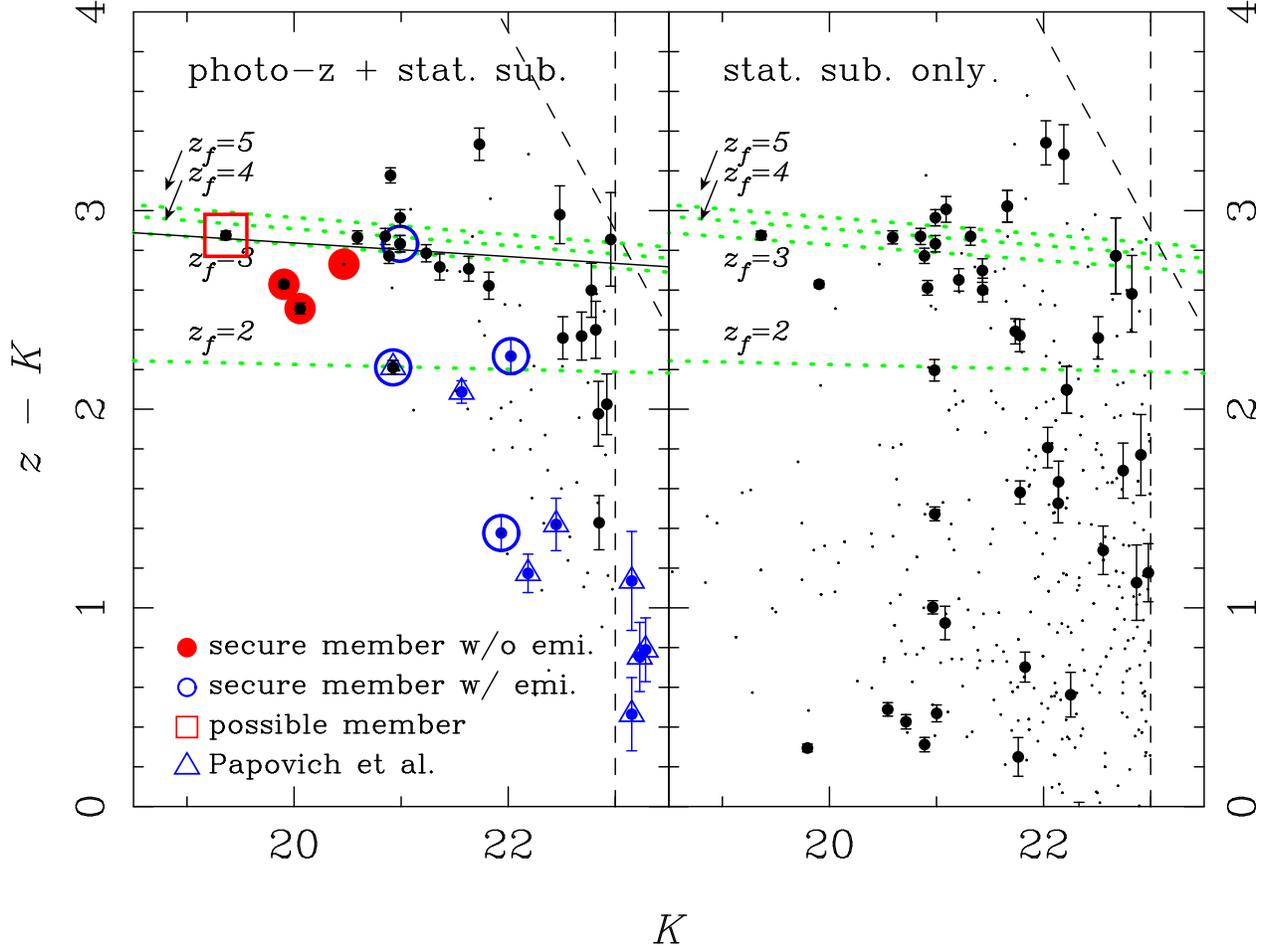}
\caption{
$z-K$ plotted against $K$.
Galaxies in the left panel are selected first by photometric redshifts
($1.4<z_{phot}<1.8$) and then the remaining fore-/background contamination
within the photo-$z$ slice is statistically subtracted.
The dot and points show subtracted and remaining galaxies, respectively.
We also show spectroscopic objects at $1.62<z_{spec}<1.65$ 
and the meanings of the symbols are shown in the panel.
The dashed lines show the $K$-band magnitude cut and $5\sigma$ limiting colors.
The dotted lines are the model red sequences formed at $z_f=2$, 3, 4, and 5 from
bottom to top.
The solid line is a biweight fit to bright red galaxies ($K<22$ and $z-K>2.5$).
We apply only the statistical field subtraction without using photo-$z$
in the right panel.
}
\label{fig:cmd}
\end{figure}

\clearpage

\begin{figure}
\epsscale{1.0}
\plotone{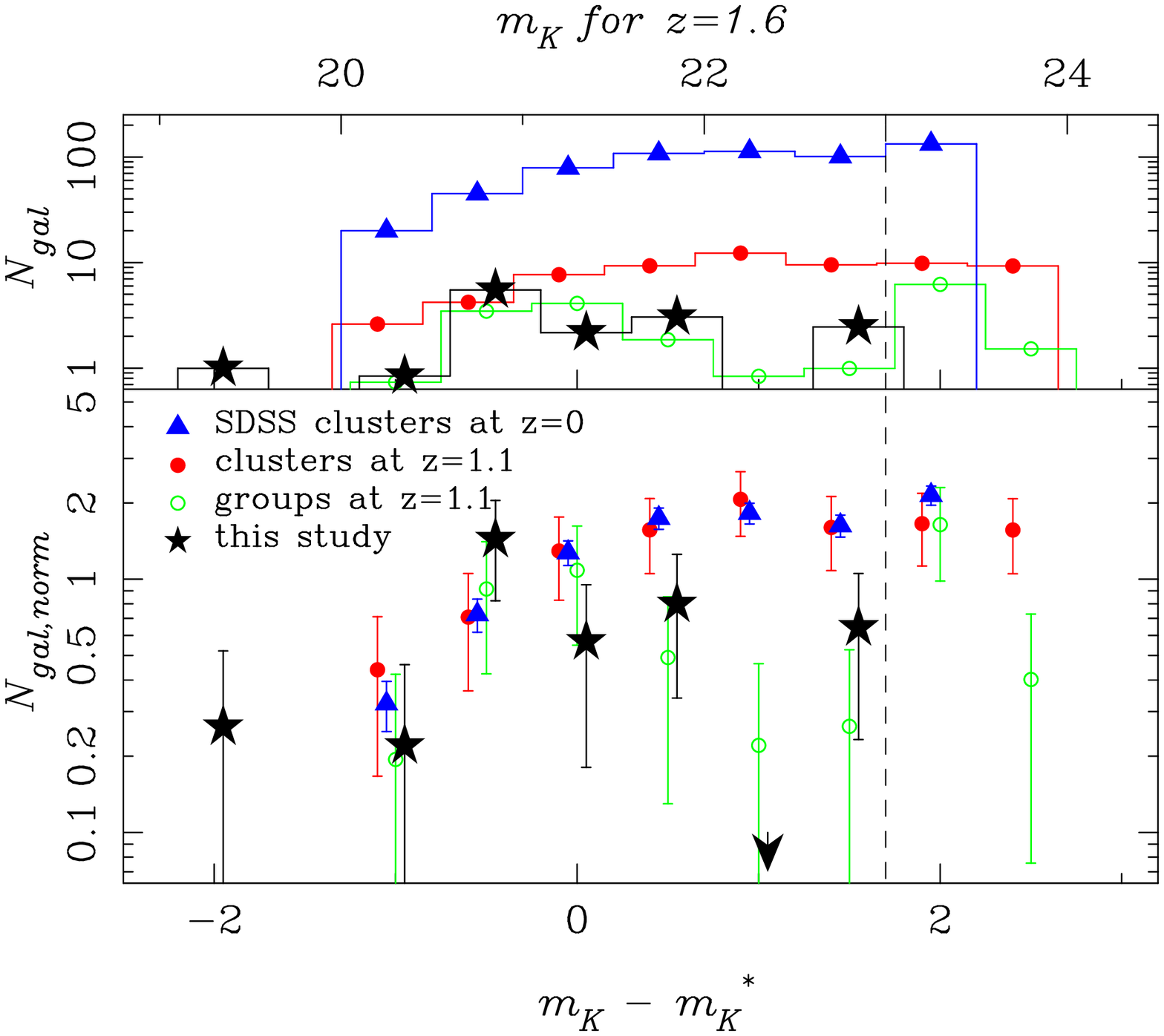}
\caption{
LFs of red galaxies.  The top panel show the $N_{gal}$
(after the field subtraction) and the bottom panel show $N_{gal}$
normalized at $\sim m_K^*$ at each redshift.
The top axis shows $K$-band magnitudes for the $z=1.6$ galaxies,
while the bottom axis shows $K$-band magnitudes relative to
$m_K^*$ at each redshift.
The meanings of the symbols are shown in the plot.
At $z=1.6$,  $m_{K}^*$ is derived by evolving $m_K^*$ at $z=0$
back in time assuming $z_f=4$.
The points are shifted horizontally to avoid overlapping.
The LFs at $z=0$ and $z=1.1$ are taken from \citet{tanaka08}.
The dashed line shows the magnitude cut of $K=23$ for the $z=1.6$ sample.
}
\label{fig:lf}
\end{figure}

\clearpage

\begin{figure}
\epsscale{1.0}
\plotone{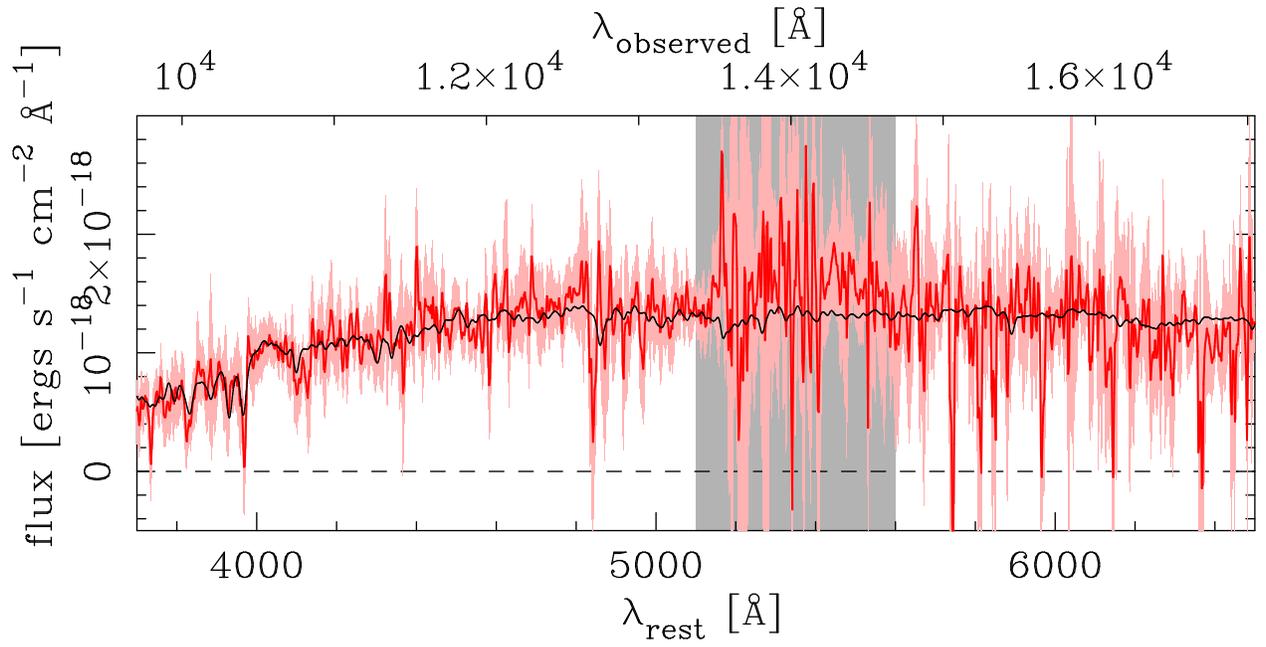}
\caption{
The observed spectrum of the brightest red galaxy with secure redshift
(dark and light coded are the spectrum and its error) overlaid with
the best-fitting model spectrum.
The shaded region is not used in the fit to avoid effects of
strong atmospheric extinction.
}
\label{fig:specfit}
\end{figure}

\end{document}